\documentclass[conference]{IEEEtran}
\IEEEoverridecommandlockouts

\newcommand{\conftitle}{15th International Workshop on Science Gateways (IWSG2023), 13-15 June 2023}

\usepackage{fancyhdr}
\usepackage{graphicx}
\usepackage{hyperref}
\fancypagestyle{pageStyle}{%
    \fancyhf{}
    \fancyhead[C]{\conftitle}

}


\usepackage{listings}
\ifCLASSINFOpdf
\else
\fi
\begin{document}

%
\title{An approach to provide serverless scientific pipelines within the context of SKA}
\renewcommand\IEEEkeywordsname{Function as a Service, Serverless, HPC, Cloud Computing, SKA, radio astronomy, Open Science}

\author{\IEEEauthorblockN{Carlos Ríos-Monje, Manuel Parra-Royón,  Javier Moldón,\\Susana Sánchez-Expósito, Julián Garrido \\Laura Darriba, MAngeles Mendoza, Jesús Sánchez and\\Lourdes Verdes-Montenegro} 
\IEEEauthorblockA{Instituto de Astrofísica de Andalucía, Gta. de la Astronomía, s/n, 18008 Granada,Spain\\
Corresponding author: mparra@iaa.es}
\and
\IEEEauthorblockN{Jesús Salgado}
\IEEEauthorblockA{SKA Observatory, \\Jodrell Bank,\\ Macclesfield SK11 9FT, \\UK}
}



%



\maketitle
\thispagestyle{pageStyle}
\pagestyle{fancy}
\renewcommand{\headrulewidth}{0pt} 

\begin{abstract}

Function-as-a-Service (FaaS) is a type of serverless computing that allows developers to write and deploy code as individual functions, which can be triggered by specific events or requests. FaaS platforms automatically manage the underlying infrastructure, scaling it up or down as needed, being highly scalable, cost-effective and offering a high level of abstraction. Prototypes being developed within the SKA Regional Center Network (SRCNet) are exploring models for data distribution, software delivery and distributed computing with the goal of moving and executing computation to where the data is. Since SKA will be the largest data producer on the planet, it will be necessary to distribute this massive volume of data to the SRCNet nodes that will serve as a hub for computing and analysis operations on the closest data. Within this context, in this work we want to validate the feasibility of designing and deploying functions and applications commonly used in radio interferometry workflows within a FaaS platform to demonstrate the value of this computing model as an alternative to explore for data processing in the distributed nodes of the SRCNet. We have analyzed several FaaS platforms and successfully deployed one of them, where we have imported several functions using two different methods: microfunctions from the CASA framework, which are written in Python code, and highly specific native applications like wsclean. Therefore, we have designed a simple catalogue that can be easily scaled to provide all the key features of FaaS in highly distributed environments using orchestrators, as well as having the ability to integrate them  with workflows or APIs. This paper contributes to the ongoing discussion of the potential of FaaS models for scientific data processing, particularly in the context of large-scale, distributed projects such as SKA.

\end{abstract}

\begin{IEEEkeywords}  
\end{IEEEkeywords}

%
\IEEEpeerreviewmaketitle

\section{Introduction}
The evolution of computing paradigms has been driven by the need to process ever-increasing amounts of data and perform complex computations more quickly and efficiently, benefiting many scientific fields. With the advent of personal computers, client-server computing emerged and scientific applications were divided into two parts: the client (user access and interface) and the server (back-end processing) where the client would request data from the server, which would process the request and return the data to the user. 

As computational and storage requirements increased, distributed processing emerged as a way to leverage the processing power of multiple computers or clusters connected via a network allowing for faster processing of large datasets and complex computations in many scientific fields. Grid computing was one of the most successful models to provide a distributed computing environment at the time but it was also a complex system. High Performance Computing uses supercomputers with large volumes of memory and specialised architectures where hardware is highly optimised for certain processes, to perform complex calculations and simulations and it is used in scientific research, engineering and other fields where large amounts of data need to be processed extremely quickly. In the early 2000s until today, as a way to provide on-demand access to computing resources over the internet, Cloud computing (CC) as computing paradigm allows users to get computing resources (such as Virtual Machines [VMs], storage, networks or software as services) on a pay-on-demand basis.

CC has brought a revolution in the way computing, storage and other resources are managed, giving institutions, businesses and individuals the ability to harness the potential in the form of very different on-demand services. CC provides flexible access to computing resources that can be scaled up or down as needed, allowing researchers to process and analyze large amounts of data efficiently as well as to provide access to specialized computing resources such as GPUs and FPGAs that can accelerate scientific pipelines. CC makes it easier for researchers to collaborate  sharing data or the execution environment, and to publish and share their work through cloud-based services, being digital infrastructures a key player for scientific progress and Open Science. Within CC, a paradigm called Serverless \cite{IEEEhowto:kopka} or Function-as-a-Service (FaaS) has emerged in recent years, which allows code to be executed (as functions) without having to manage \cite{comps} or provision VMs, computing power or storage resources. Serverless allows you to focus on writing and deploying code or functions, being this is a natural extension of CC, as it provides even more flexibility and scalability than traditional CC by allowing developers to split their applications/services into smaller, more manageable functions that can run independently and in a more streamlined way. 

The Square Kilometre Array (SKA) is a new generation radio telescope currently under construction and will be one of the most advanced and powerful telescopes in the world, capable of observing the sky with an unprecedented level of detail and sensitivity. It will deliver about 700 PB of data products per year. Given these attributes, the SKA will require significant computing resources for both data processing and analysis. The SKA data will be delivered to a Global Network of SKA Regional Centres (SRCNet) which will provide access for an international community to SKA Observatory data and the analysis tools as well as the processing power to fully exploit their science potential, so SRCs is where the SKA science will be done.  It will require the use of advanced computing technologies and techniques, such as HPC, distributed CC and Machine Learning (ML). 

In this context, we believe that Serverless computing can be highly beneficial to SRCNet data processing capabilities due to its scalability, cost-effectiveness, efficiency and reliability. With Serverless, researchers can scale their computing resources as needed and ensuring that SKA data processing workflows are highly available across different global regions, as well as they can design functions that can be integrated from anywhere, such as Jupyter Notebooks, workflows or command line, among others. In this paper we explore the feasibility of using Serverless computing for the design of several functions commonly used in astronomy pipelines and in particular of some of the operations involved in the data workflows of SKA precursors telescopes (such as MeerKAT, ASKAP or MWA) and to tackle image analysis, data cube visualization and analysis, spectral analysis, source extraction, among others.  

As for the content, in the section \ref{sec:background} we will review the application of this Serverless paradigm to scientific problems, then in section \ref{sec:approach} the different solutions available to deal with the model from the point of view of architecture, services and composition will be addressed as well as we will detail the serverless and FaaS approach that has been chosen including a subset of radio interferometry functions that have been implemented. Finally, in section \ref{sec:conclusions} we address the conclusions and the future work that we propose.

\section{Background}
\label{sec:background}

As commented, Serverless computing is a popular CC model that allows developers to run code and applications without worrying about underlying infrastructure. This model provides scalability and cost-effectiveness by automatically handling the provisioning and management of servers, scaling resources based on demand, and handling fault tolerance and availability. Cloud providers, such as Amazon Web Services, Microsoft Azure, and Google Cloud, offer serverless computing services that provide a wide range of capabilities \cite{marketplace}, from simple function execution to complex event-driven architectures, and integrate with other cloud services like storage, network or more recently the use of ML or Artificial Intelligence (AI) capabilities as a service \cite{mlserverless}.

Serverless computing is being adopted across a wide range of industries \cite{serverlessindustry} and use cases, from web and mobile applications to Big Data processing and scientific computing. However, there are still challenges to overcome, such as the lack of standardized development and deployment practices, limited tooling and debugging capabilities, and issues related to security and compliance. Despite these challenges, the state-of-the-art in serverless computing is rapidly evolving, with new features and services being added to cloud provider offerings \cite{serverlessedge}, and the adoption of serverless computing increasing among developers and organizations.

In the CC context, the current serverless landscape was introduced during an Amazon Web Services 'Reinvent' event in 2014 , and since then, multiple cloud providers and industrial and academic institutions have introduced their own serverless platforms. Serverless platforms are a natural step after VMs and container technologies, where each stage led to lighter computing units in terms of resource consumption, cost and development speed. Other CC providers followed in 2016 with the introduction of Google Cloud Functions, Microsoft Azure Functions and IBM OpenWhisk, the latter released as Open Source (OS) to the community. In addition to public cloud providers that support FaaS, there are several initiatives to bring Serverless to private environments from OS developments ready to be deployed in our own infrastructures, such as OpenFaaS or OpenLambda, among others. FaaS has been enriched with high-performance, lightweight fast polyglot runtimes, and Ahead-of-Time (AOT) compilation features for a faster startup times and lower memory usage, as proposed with GraalVM\footnote{\url{https://www.graalvm.org/latest/docs/introduction/}}.

Within the scientific environment there are several research works on the application of the serverless model to workflows in different areas of knowledge. In \cite{Kijak}, authors look into the problem of scheduling scientific workflows and discussing challenges related to workflow scheduling with Cloud functions. Other studies \cite{Malawski} address FaaS execution systems with a hybrid solution for scientific workflows and from another point of view in \cite{Lap}, challenges in serverless edge computing and open research opportunities for ML applications are depicted.  Thus, within the field of astrophysics, the serverless model has still not been sufficiently tackled. In this paper we want to initially explore how FaaS can be an option for the deployment and execution of scientific workflows in this field of knowledge.


\section{Serverless computing approach for workflows in astronomy}
\label{sec:approach}


Serverless computing is typically implemented using a Function-as-a-Service model. With FaaS, developers and scientists write small, self-contained functions that perform a specific task or respond to a particular event. These functions are deployed to a serverless platform (such as AWS Lambda, Google Cloud Functions, or OS FaaS platforms in your own infrastructures) and are triggered by events, such as HTTP requests, database updates, or messages from a queue collector. When a function is triggered, the system will automatically allocate the resources regardless of the underlying platform, either HPC, CC, etc. to run the function. In this case the function runs in a stateless container, meaning that it doesn't store any persistent state or data between invocations. Once the function completes its task, the container is terminated, and the resources are released back to the platform.

\begin{figure}[t]
\includegraphics[width=\columnwidth]{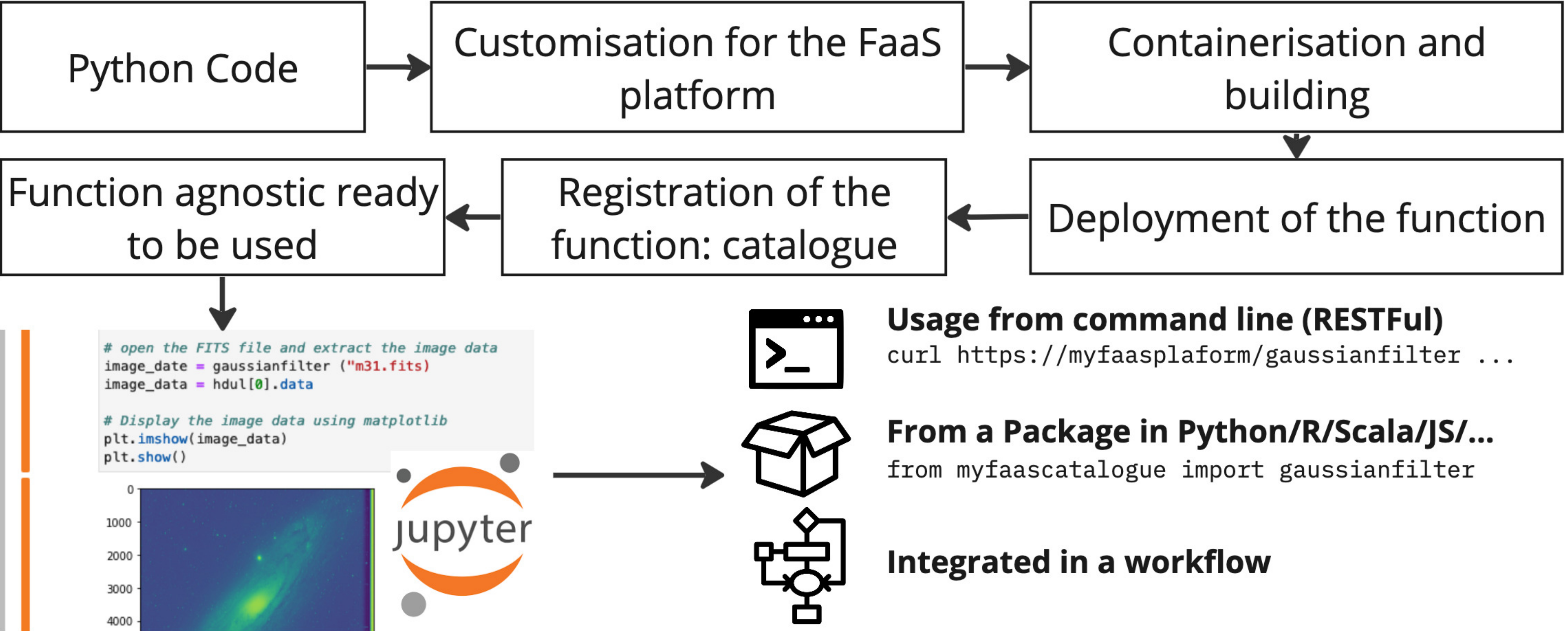}
\centering
\caption{Steps from a simple python code to a function in a FaaS platform.} 
\label{fig:serverless-steps}
\end{figure}

In this scenario if we want, for example, to create a function that takes an image, applies a Gaussian blur filter to smooth the image and returns the image data, we just need to write the code as usual and add some handlers for the input and output data of the function as shown in code listing \ref{lst:listinggauss}. Once this is done, it will be necessary to build the function using a language such as Python, Go, Scala, NodeJS, or others. This code is then included into a container that provides the core component to provide the FaaS integration logic. The containerised function must then be registered in the catalogue within the FaaS platform so that it is available to be invoked from HTTP or from an API. Finally, the function can be called from virtually any application or service as it is published from a RESTful API or GraphQL, for example. With this publishing model the function could be used from a Jupyter Notebook for example as shown in the Figure \ref{fig:serverless-steps}.

\begin{lstlisting}[caption={Example of an image filtering for a FaaS deployment.},label={lst:listinggauss},language=Python]
import cv2
import numpy as np
import astropy.io.fits as fits
from flask import Flask, request, Response
def main():
    # get the file from the request
    file = request.files['file']    
    # read the FITS image
    hdul = fits.open(file)
    image = hdul[0].data    
    # Gaussian filter:kernel 3x3, sigma 1.5
    fi = cv2.GaussianBlur(image, (3, 3), 1.5)    
    # write FITS file
    hdul[0].data = fi
    hdul.writeto(file, overwrite=True)    
    # return the filtered FITS
    return Response(open(file, 'rb').read(), 
       mimetype='application/fits')
\end{lstlisting}


Public serverless computing platforms, such as AWS Lambda or Google Cloud Functions, are offered as fully managed services by CC providers, meaning that the it takes care of all aspects of managing and scaling the platform, including underlying infrastructure, security, and maintenance as well as to provide a convenient and scalable way for developers to build and deploy serverless applications without having to manage the underlying infrastructure and paying-as-you-go. On the other hand, OS serverless computing platforms, are community-driven projects that provide an alternative to commercial services. These platforms are often built on top of container orchestration systems like Kubernetes or Docker, and allow developers to run serverless functions on their own infrastructure or in a public cloud.

Open Science has been gaining momentum in recent years, and many research organizations are looking for ways to adopt it into their workflows. One of the key components of Open Science is the ability to share research data and code openly and transparently. This is where OS platforms for Serverless computing and Function-as-a-Service (FaaS) capabilities can be of great value. In order to make it easy for research organizations to adopt these platforms, it's important to provide here not only the code to run the platform and create the containers, but also the logic and software for the functions implemented on the platform. This ensures that researchers have a deep understanding of what the functions are doing internally, and can customize them to fit their specific needs, leading to more efficient and effective research. In this paper we have provided a repository (see section \ref{sec:approach}) with all these components to promote Open Science and making it easier for research organizations to adopt FaaS capabilities.


In table \ref{tab:faasplatformsstudied}, we provide with summary of these platforms that we have been working with to develop the set of functions detailed in subsection \ref{subsec:ourworkflow}. 

\begin{table*}[]
    \centering
\begin{tabular}{|c|c|c|c|c|}
\hline
Platform & State of development & Difficulty of installation & Cluster & Supported Languages \\ \hline\hline
OpenFaaS & Active   & Easy  & Kubernetes or OpenShift & Go, Python, Node.js, Ruby, Java, C\#, Dockerfile, etc.\\ \hline
Knative  & Active  & Moderate & Kubernetes  & Node.js, Python, Go, Quarkus, etc \\ \hline
Kubeless &  Not actively maintained & Easy & Kubernetes & Go, Python, NodeJS, Ruby, etc. \\ \hline
Fission  & Active & Easy & Kubernetes or OpenShift     & Node.js, Python, Ruby, Go, PHP, Bash and Dockerfile \\ \hline
OpenWhisk & Active &  Moderate     & Kubernetes     & Node.js, Python, Swift, Java and more \\ \hline
\end{tabular}
\caption{Open Source FaaS Platforms}
\label{tab:faasplatformsstudied}
\end{table*}


\subsection{Serverless with an astronomy workflow}
\label{subsec:ourworkflow}


Within the context of working with and processing data from both SKA precursors and pathfinders telescopes\footnote{\url{https://www.skao.int/en/explore/precursors-pathfinders}}  as well as other radio telescopes, the analysis of interferometric data involves the concatenation of a series of steps, generally in linear way, but which can be approached from different points of view, being organised in a logical chain or workflow with/without dependencies. The steps broadly include data manipulation, where raw data from the radio telescope are managed to ensure that they are in a suitable format for analysis and all necessary conversions are performed. This is followed by flagging, which involves the removal of incorrect or corrupted data or data affected by noise or other artefacts, to ensure correct calibration and imaging of the data can be obtained. The next step would be calibration, which aims to generate calibration tables and apply them to the data to correct for instrumental and atmospheric effects. Usually, the data of the calibrated target source is splitted and averaged if needed. And finally the imaging phase and self-calibration, where deconvolution, cleaning,  imaging and residual calibration is performed for a final product of images or data cubes. Note that some instruments or observing modes will require variations of this general workflow, in particular with additional steps and verification functions (see an overall workflow in \cite{wf}.


To perform these steps, several astronomical tools can be used. The most widely used library package for interferometric data analysis is CASA \cite{adass01}, currently implemented as Python modules, so it can be run interactively or via scripts or workflows. There are also several Python-based pipelines for different instruments, such as VLA CASA calibration pipeline pipeline, the e-MERLN CASA pipeline \cite{emerlin}, CARACAL \cite{caracal}, among others, which include tools and containers customised with the pipeline operations. Additionally, there are different standalone tools (binary executables) such as wsclean \cite{wsclean} (for imaging), aoflagger \cite{aoflagger} (flagging) that are specialised in specific aspects of the pipeline and can be much more efficient than CASA on specific tasks, including support for GPUs in some cases. In this way each step from the beginning of the raw data to the presentation of the image is a succession of steps where the results of each step are the inputs to the next ones, so that the processing can be highly automated and if possible distributed and scaled.

Under this view the workflow would typically involve the development of individual functions or microservices that perform specific tasks, such as data manipulation, cleaning, flagging, calibration or/and image processing. These functions would then be orchestrated together, selecting the most appropriate tool for each relevant combination of instrument and scientific objective, to form the overall serverless application by using a FaaS platform. This model offers advantages in terms of high availability, scalability and interoperability of tasks in a highly distributed environment such as the SRCNet, with globally distributed centres providing different computing platforms and diverse hardware, as well as configuring a SRCNet datamesh model\footnote{\url{https://www.datamesh-architecture.com/}}. FaaS would use an "execution planner" service to determine the optimal location for executing the function based on the location of the data, available resources, and cost. The function would then be deployed to the appropriate data center, minimising large data transfers and latency and maximising performance.


In a simple context, from a simple FaaS model, functions are called on demand with no \textit{a priori} scheduling on where or how they will be executed, as they are abstracted for the user, and it is the underlying container orchestrator that manages it with its policies. But in a complex and highly distributed environment such as a datamesh, it is necessary to operate the functions with an execution planner at the orchestrator level that transfers/decides the execution/computation of these functions to the specific resources where they are best suited. Faas can help this model from the orchestrator by analysing the data and determining which functions should run on which parts of the data, distributing the computational workload across the distributed network to make decisions based on factors such as data availability, network latency and cost, ensuring that functions run in the most efficient and cost-effective way.



\subsection{A testbed platform for serverless with Fission}
\label{subsec:fission}

We have studied and tested Fission (see table \ref{tab:faasplatformsstudied}) as an OS platform that could be applied to the scientific analysis to produce SKA advance data products within the SRCNet platform. Fission has become more popular in the recent times 
due to aspects such as ease of use, implementation and deployment, integration with Kubernetes natively and scaling of functions using Kubernetes metrics, as well as the need to reduce the latency of functions in highly distributed environments, providing a fast cold start time of functions. 

For this test bed we have deployed Fission on the Spanish Prototype of an SKA Regional Centre, the SPSRC \cite{towards}, located in Granada, Spain, under a two-node Kubernetes cluster (v.1.24.0, using Rancher as orchestrator and Fission - release 1.6.0) and exposing the functions locally to the SPSRC projects. The installation steps can be found in the next section.


\subsection{Development and deployment of functions}
\label{subsec:develdeployfunctions}

Based on the radio interferometry workflow, we have selected a subset of functions that can be easily exported within the FaaS Fission platform. To show the adaptability of FaaS to practically any code/engineering or application that the user wants to execute, we will show as an example how to design a function for the imaging component \textit{tclean} from CASA framework and as well the \textit{wsclean} application. With these two possibilities, Python CASA code and binary application, we can highlight the capabilities of FaaS to be extended to any kind of software, as they can be easily containerised. 

For reasons of space, all the details of installation, environments, execution and parameterization are available from the project repository\footnote{\url{https://github.com/manuparra/ska-serverless}}.

The first step towards creating a function in Fission is to define the environment where the function works. This environment will need all the necessary dependencies and packages, in addition to the FaaS Fission core components. We can create a custom environment from one of the pre-existing Fission environments\footnote{Fission environments: \url{https://github.com/fission/environments}}. To use CASA framework, we will need to customise one Python environment with the corresponding Python packages. For the first approach, starting from a Python environment, we modify the Dockerfile, as indicated in \ref{lst:dockerfiletclean}, to choose a proper base image, \textit{python:3.8-buster} and then, we add the CASA packages, \textit{casatasks}, \textit{casatools} and \textit{casadata}, to the \textit{requirements.txt} file. In code listing \ref{lst:dockerfiletclean} the last two lines include the basic services for FaaS to run on the platform.  Then, we build the image and push it to a public repository like DockerHub. With this we have the image fixed with the Fission core services and all the dependencies, ready to create functions that will include the CASA framework.

\begin{lstlisting}[caption={},label={lst:dockerfiletclean},language=bash,caption={Dockerfile for a customised environment with python and CASA.}]
FROM python:3.8-buster
RUN apt-get update -y && 
    apt-get install -y python3-dev libev-dev
WORKDIR /app
COPY requirements.txt /app
RUN pip3 install -r requirements.txt
COPY . /app
ENV PYTHONUNBUFFERED 1
ENTRYPOINT ["python3"]
CMD ["server.py"]
\end{lstlisting}

The second approach is an example with wsclean a binary imaging application. We need to use another template that provides support for this kind of environment. When it comes to creating FaaS functions that call specific applications, it will be necessary that the Dockerfile\footnote{Customised environment for wsclean: \url{https://github.com/manuparra/ska-serverless/tree/main/Fission/enviroments/wsclean}} contains the specific installation of the software in the version we want to use. For this case, we prepared an environment for wsclean 3.3 with support for EveryBeam, Dysco and IDG. Then we build the image and upload it to an image repository such as DockerHub.

\begin{lstlisting}[caption={},label={lst:fissionenvs},language=Bash,caption={Adding FaaS enviroments for CASA and wsclean.}]
fission environment create --name \
    python-casa-6.5 --image dockerhub/casa 
fission environment create --name \
    wsclean-3.3 --image dockerhub/wsclean
\end{lstlisting}

Once we have the containers built, the next step is to create an environment within Fission. This environment will allow Fission to know how to manage the logic of the function we are going to implement. To do this, just use the Fission CLI and execute both commands from the \ref{lst:fissionenvs} list. The first one enables an environment for the functions with CASA and the second one with wsclean.

With both environments in place, it is now time to create the logic for the functions in the Fission platform. The creation of the function consists of developing a code in the language selected in the environment (Python and native application in our case), which includes the parameterisation and the starting point from where our function will start executing. For the environment with Python and CASA, we are going to design a pipeline operation to allow radio interferometric image reconstruction, using \textit{tclean}. Code listing \ref{lst:fissioncasa} shows how to capture the input parameters with \texttt{request.get\_json()}, such as the input/output data and the execution parameters, then we define the function to be executed in particular from CASA  \texttt{ct.tclean} and finally the function returns where the output data is stored.

\begin{lstlisting}[label={lst:fissioncasa},language=python,caption={FaaS Fission function for one operation with tclean.}]
from flask import request
import os, json
import casatasks as ct
def main():
    param = request.get_json()
    input = "/data/" + param["Input-MS"]
    output = "/data/" + param["Output-MS"]    
    ct.tclean(vis = input, 
              imagename= output, 
              **param)
    return "/data/" + output
\end{lstlisting}

For the design of the function with \texttt{wsclean}, it can be done natively using a bash code to execute the binary command, but it can also be integrated and called from Python code using \texttt{subprocess}. Both options are perfectly compatible. In our case, for convenience in the interfaces, we will use a wrapper from Python to call \texttt{wsclean} as indicated in code listing \ref{lst:fissionwsclean}.

\begin{lstlisting}[label={lst:fissionwsclean},language=python,caption={FaaS Fission function for one operation with wsclean.}]
from flask import request
import os, json
def main():
    param = request.get_json()
    input = "/data/" + param["Input-MS"]
    parameters_str = ... # Extract parameters
    subprocess.run(["wsclean"] + 
        parameters_str.split() + [input])
    return "/data/" + output
\end{lstlisting}

For this testbed, we've use an already prepared data that can be downloaded from CASA repository\footnote{\url{https://casaguides.nrao.edu/index.php?title=VLA_CASA_Imaging-CASA6.2.0}}. 

Finally, one more operation must be carried out, which consists of adding the function to the platform and publishing it in the catalog. In the code listing \ref{lst:fissionfns}, our functions are integrated and published in two URLs to be able to consume each function independently, in this case in our server \texttt{https://server/function/} where \texttt{function} is \texttt{tclean} or \texttt{wsclean}.

\begin{lstlisting}[label={lst:fissionfns},language=Bash,caption={Two functions created and published in Fission.}]
fission fn create --name tclean --env python-casa \
  --code tclean.py --method POST --url "/tclean/" 
fission fn create --name wsclean --env wsclean \
  --code wsclean.py --method POST --url "/wsclean/" 
\end{lstlisting}

To perform some tests, we can call the functions created from the command line by accessing the URL, previously generated and sending it the data and parameters, as indicated in the code listing \ref{lst:curl}.

\begin{lstlisting}[label={lst:curl},language=Bash,caption={Test tclean function with data and parameters.}]
curl -X POST -d "$(cat parameters-tclean.json)" \
   -H "Content-Type: application/json" \
   "http://${OURFaaSPlatform}/tclean/"
\end{lstlisting}

With all of this procedure, it is possible to expand the function catalog with virtually any type of programming language, application or container so that through the underlying orchestrator, these functions can be scaled indefinitely and interoperated  from a workflow or API. The scientific user can publish their functions in source code and containers from a public repository so that they can be shared within the community and then integrated into FaaS.

In terms of metrics, we have tested these functions in a basic way with a small dataset example, obtaining consistent times with a minimal overhead due to management on the orchestrator. These executions can be found in the project repository, and this study is proposed as part of future work where it will include benchmarks to showcase performance in more complex and realistic working environments.

\section{Conclusions and future work}
\label{sec:conclusions}

The SKA, currently under construction, will be the largest radio interferometer, and will be the largest producer of public data on Earth. The SKA Regional Centre will be a federated network of research and computing facilities dedicated to enable researchers to convert SKA data to advanced data products and finally to scientific results. This data processing challenge requires innovative, efficient and robust systems able to scale the workflows while giving enough flexibility to adapt to the complexity of the data products and the necessities of specific research programs. In this context,
as an advantage over using containers, FaaS abstracts away infrastructure management and scalability, provides greater portability and flexibility by allowing functions to run on different platforms and environments without additional changes.
We have been aiming to implement a platform for FaaS and create different operations of a radio astronomy workflow as functions available in a catalogue that can be called from virtually any API, service or library, so that we can abstract both the software that runs internally in the function and the scaling of the computational resources it uses. We have encapsulated existing code/libraries and applications within containers to be distributed globally through a deployment of federated orchestrators, with the aim of ensure the high availability and efficiency of functions executed for example on a large scale as is the SRCNet. This approach is in line with the model of moving computation to where data are located, resulting in reduced latency and improved overall performance. 

FaaS still needs to be explored in detail to understand its constraints such as, per-function memory scaling, capabilities similar to AWS SnapStart, and proxy services for low-latency access, databases or data/computing co-location. Therefore, as future work, we propose different aspects to be worked on within the context of SRCNet. On the one hand, it would be interesting to test the performance of the functions developed to monitor performance with data volumes of similar orders of magnitude to those available in SRCNet. Additionally, on the other hand, we propose studying how to implement an "execution planner" that can deliver functions as close as possible to the data, as a key element for SRCNet.

\section*{Acknowledgment}

The authors acknowledge the Spanish Prototype of an SRC (SPSRC) service and support funded by the Spanish Ministry of Science, Innovation and Universities, by the Regional Government of Andalusia, by the European Regional Development Funds and by the European Union NextGenerationEU/PRTR. The SPSRC acknowledges financial support from the State Agency for Research of the Spanish MCIU through the "Center of Excellence Severo Ochoa" award to the Instituto de Astrofísica de Andalucía (SEV-2017-0709) and from the grant CEX2021-001131-S funded by MCIN/AEI/ 10.13039/501100011033. Carlos Rios acknowledges Scholarships of introduction to research, "JAE Intro ICU" funded by CSIC, "Programa JAE". We acknowledge financial support from the grant  PID2021-123930OB-C21 funded by MICIU/AEI/ 10.13039/501100011033 and by ERDF/EU. We also acknowledge financial support from the grant TED2021-130231B-I00 funded by MICIU/AEI/ 10.13039/501100011033 and by the European Union NextGenerationEU/PRTR.



%


\end{document}